\documentclass[11pt,en,cite=authoryear]{elegantpaper}

\title{Unification of global height system at centimeter level using precise frequency signal links}
\author{Ziyu Shen$^1$  \and Wen-Bin Shen$^{2,3,*}$ \and Shuangxi Zhang$^{2,3}$}
\institute{$^1$ School of Resource, Environmental Science and Engineering, Hubei University of Science and Technology, Xianning, Hubei, China \\
$^2$ Time and Frequency Geodesy Research Center, Department of Geophysics, School of Geodesy and Geomatics, Wuhan University, Wuhan, China  \\
$^3$ Key Lab of Surveying Eng. and Remote Sensing, Wuhan University, Wuhan, China \\
$^*$ Corresponding author: wbshen@sgg.whu.edu.cn}

\usepackage{array}  
\usepackage{amsmath,bm}     
\usepackage{url}

\begin{document}

\maketitle

\begin{abstract}
The realization of International Height Reference System (IHRS) is one of the major tasks of the International Association of Geodesy (IAG). 
A main component of the IHRS realization is the global vertical datum unification, which requires the connection of the existing local vertical height reference systems (VHS).
However, it is difficult to estimate the offsets between two local height systems by conventional approaches when they are far apart.
In this paper, we formulate a framework for connecting two local VHSs using ultra-precise frequency signal transmission links between satellites and ground stations, which is referred to as satellite frequency signal transmission (SFST) approach.
The SFST approach can directly determine the geopotential difference between two ground datum stations without location restrictions, and consequently determine the height difference of the two VHSs.
Simulation results show that the China's VHS and the US's VHS can be unified at the accuracy of several centimeters, provided that the stability of atomic clocks used on board the satellite and on ground datum  stations reach the highest level of current technology, about $4.8\times 10^{-17}\tau^{-1/2}$ for an averaging time $\tau$ (in seconds).
The SFST approach is promising to unify the global vertical height datum in centimeter level in future, and it also provide a new way for the IHRS realization.
\end{abstract}

\section{Introduction}
Reference frames with long-term stability and homogeneous consistency worldwide plays a key role in establishing various theoretical frameworks, e.g., gravity field, Earth rotation, geodynamics, as well as extensive applications, such as global navigation satellite system, precise positioning, precise observations of any subject in space, etc. 
The International Terrestrial Reference System (ITRS) and its realization, International Terrestrial Reference Frame \citep[ITRF, ][]{Petit2010-hu} provide a globally unified geometric reference frame with accuracy at millimeter level. 
However, currently an equivalent high-precise global physical reference frame that reflects the Earth’s gravity field is still missing \citep{Ihde2017-wk}. 
In order to establish a consistent and accurate physical reference frame, the International Association of Geodesy (IAG) released the IAG Resolution No.1. for the definition and realization of an International Height Reference System (IHRS) in 2015 \citep{Sanchez2016-nd}, which design a physical world height system as the basis for monitoring effects generated by gravity field variation. 
Similar to the geometric reference system and frame, the realization of IHRS is the establishment of the International Height Reference Frame (IHRF).
The IHRS is defined by an equipotential surface of the Earth's gravity field, where the geopotential value at the surface is the conventional value $W_0=62 636 853.4 m^2/s^2$ (zero-height level), and the vertical coordinates are geopotential numbers $C_p = W_0 - W_P$ \citep{Sanchez2016-nd,Ihde2017-wk}.
A key concept of realizing IHRS is the unification of local vertical height systems (VHSs, which refer to local isolated level) around the world, and connect them to the global one. 
Since local VHSs are usually based on mean sea level (MSL) determined by tide gauges, and the MSL is not an equi-qeopotential surface, different local VHSs exhibit inconsistencies with respect to each other up to $1\sim 2$ meters \citep{Sideris2015-fn}.
How to find out the offset between arbitrary two different height systems' origins (datums) is the main challenge for realizing IHRS, and various approaches have been tested and discussed.


Currently there are four approaches that are extensively discussed and practically applied for the height system unification. They are briefly explained as follows, each of which presents advantages and drawbacks.

(1) The conventional approach is leveling with gravity reductions. 
This is mainly used for the realization of local VHS and the accuracy can reach sub-millimeter level between neighboring leveling point \citep{Sanchez2017-zl}. 
However, leveling is laborious and time-consuming while the errors accumulate over long distances. 
In addition, the main drawback of leveling is that it cannot connect two continents separated by the ocean, which makes it impractical for the realization of a global VHS \citep{Ihde2017-wk}.

(2) Oceanic leveling \citep{Stocker-Meier1990-kj}, in contrary, is suitable for connecting different local height systems separated by oceans. 
For example, ocean models can provide a mean dynamic topography correction to height datums of countries with coastlines, thus realize the unification.
Though the uncertainty of oceanographic modeling method can be better than a decimeter \citep{Woodworth2012-th}, 
the ocean leveling is limited to height datums near a coastlines, and for high precision it requires years of continuous observation data of tide gauges with adequate density distribution \citep{Woodworth2012-th}, which is unavailable in many places such as Africa areas.

(3) The third method is estimating the anomalous potential by solving the geodetic boundary value problem (GBVP) \citep{Rummel1988-qo}. 
 It can provide a global solution for height unification, and the precision in well-surveyed regions reaches several centimeters \citep{Gerlach2013-tw,Rangelova2016-gj,Amjadiparvar2016-lg}.
But in sparsely surveyed regions the precision drops to decimeter level \citep{Sanchez2017-zl}.
Another drawback of GBVP method {lies in that it requires a prior information} of potential or height values from various sources (global geopotential model, tide gauge data, gravity observation data, et al.); {and the errors in these a prior information} will influence the precision of GBVP method, and the use of different kinds of a prior information in different regions makes it difficult to unify the height datums in these regions.

(4) The fourth method is applying global gravity models (GGMs) with high precision.
The EGM2008, for example, is complete with degree and order of spherical harmonics up to 2159 \citep{Pavlis2012-fw}, and we can directly compute the potential $C(P)$ of any given point in the ITRF coordinates by introducing it into the spherical harmonic expansion equation.
However, at present the GGMs method meets the problem of precision and resolution trade-off.
For instance, the GOCE series models (see e.g., \cite{Hirt2012-vy}; also see the released products from ESA (\url{www.esa.int}) and the International Centre for Global Earth Models (\url{icgem.gfz-potsdam.de/ICGEM})) can reach the accuracy of 1 cm (even higher) but with poor resolution of $1^\circ \times 1^\circ$.
In contrary, although the EGM2008 model has a relatively high resolution of $5'\times 5'$, its average accuracy is only about 10 to 20 cm \citep{Pavlis2012-fw}. 
Another drawbacks of the GGMs method lies in that different models usually give rise to quite obvious discrepancies, because of different standards and conventions used.


Currently it is difficult to establish IHRF with high precision by any of the approaches described above.
In order to get out of the difficulties, another method, relativistic geodetic method, has gained an increasing number of attention and discussion.
The relativistic geodetic method is based on the general theory of relativity \citep{Einstein1915-tv}: precise cocks at positions at different geopotentials run at different rates.
Therefore geopotentials or geopotential difference between arbitrary two stations can be measured by precise clocks, and the corresponding height propagation based on this method is referred to as ``chronometric leveling''  \citep{Vermeer1983-lh,Bjerhammar1985-bd}.
Since relativistic geodetic method requires ultra-high precise clocks (e.g., for the precision of 1 cm, the stability of clocks should reach $1\times 10^{-18}$), it  was not payed attention for the purpose in practical applications for a long time because of the limit of clock precision.
However, with the fast development of high-precision clock manufacturing technology in recent years, the optical-atomic clocks (OACs) with uncertainty and accuracy around $1\times 10^{-18}$ and even higher level have been generated in various laboratories \citep{Mehlstaubler2018-da,McGrew2018-og,Huang2019-ez,Oelker2019-wm}. That guarantees the feasibility of actual applications of the relativistic geodetic methods. Consequently, more and more scientists pay great attention to various potential applications of the relativistic geodetic methods \citep{Muller2008-pc,Kopeikin2011-bo,Flury2016-bw,Puetzfeld2019-sl}. 

In order to compare clocks in different places, the most precise method is to connect them via optical fibre link (OFL) \citep{Riehle2017-yd}. 
Thereby an increasing number of discussions and experiments on clocks connected by OFL have been carried out and discussed \citep{Lisdat2016-pz,Takano2016-dr,Lion2017-pl,Shen2019-od}.
Recently the most precise measurement in OFL chronometric leveling is conducted by \citet{Grotti2018-ap}, who use transportable optical clocks {with uncertainties around $5\times 10^{-17}$ to determine the geopotential difference between two points in} a mountain area between France and {Italy}. 
 Though their experiments show a height discrepancy of around 20 cm between the OFL observed result and that determined by  conventional approach (leveling and gravity measurement),  the $1 \sigma$   uncertainties is limited to around 17 m \citep{Grotti2018-ap}. 
In addition, \citet{Wu2019-wq} proposed a method to unify several local height systems by clock networks connected by OFLs.
According to their simulation results, the height systems of West European region can be unified at a precision better than 1 cm, under the assumption that the clock frequency uncertainty is $1\times 10^{-18}$ .
Although {relativistic geodetic methods are} now practical and can reach high precision, the adoption of OFL limits its development. 
That is because the cost for optical fibers will increase rapidly as the distance between two clocks increases, or as the number of stations in a network increases.

Alternatively, we can compare two clocks by microwave frequency links in space, 
and even if the two clocks are not inter-visible, they can be abridged by a satellite, and the geopotential difference between them can be measured \citep{Shen1993-rb}.
This method is regarded as satellite frequency signal transmission (SFST) approach, which is detailedly discussed in \cite{Shen2016-lc,Shen2017-kg}.
Given the assumption that the stability of OACs is $1\times 10^{-18}$ within an hour, simulation experiments show that the precision of  geopotential difference between two stations on ground can reach several centimeters in height \citep{Shen2017-kg}.
Although its precision is slightly lower than that of the OFL approach, the SFST approach is much more convenient and cost much less, and it is promising for the global VHS unification and the IHRS realization.

In this paper, we propose an approach for unifying global VHS by providing example how to connect arbitrary two different local VHSs using SFST approach. 
In section \ref{sec:basic} we briefly describe the concept of height reference system and SFST method. 
In section \ref{sec:SE} we conduct simulation experiments of VHS unification by a SFST network, and present our results. 
In the last section we provide discussion and conclusions about this work and potential improvements and applications in future.

\section{Height reference system and the SFST method}
\label{sec:basic}
\subsection{International height reference system and vertical height reference system}
\label{ssec:IHRS}

The International Height Reference System (IHRS) is a geopotential reference system co-rotating with the Earth, as defined by the International Association of Geodesy {(IAG)} in 2015 \citep{Drewes2016-lc}.
According to IHRS, the geopotential on the geoid (simply geoidal potential) is a constant value $W_0 = 62 636 853.4$ $\rm{m}^2/\rm{s}^{-2}$, and the vertical coordinates are defined as \citep{Ihde2017-wk} 
\begin{equation}
    C_P = -\Delta W_P = W_0 - W_P ,
    \label{eq:C}
\end{equation}
where $C_P$ is denoted as geopotential number, $\Delta W_P$ is the geopotential difference between the potential $W_P$ at the considered point $P$ and the geoidal potential $W_0$.

A vertical height reference system (VHS) is defined by geographic elevation or depth in relation to a reference surface (which is usually the local sea level) \citep{Luz2002-nq,Sanchez2007-ac,Ihde2008-iu}.
It has close connection to the concept of IHRS because its reference surface could be the geoidal surface with $W_0$, and the value $C_P$ (given in $\rm{m}^2/\rm{s}^{-2}$) can be converted to a physical height $H_P$ (given in m) by the following equation \citep{Hofmann-Wellenhof2005-ur,Torge2012-xq}
\begin{equation}
    H_P = \frac{C_P}{\hat g} = \frac{W_0 - W_P}{\hat g} ,
    \label{eq:H}
\end{equation}
where $H_P$ can be orthometric height (OH), normal height or dynamic height, depending on the types of $\hat g$ {applied}. 

\begin{figure}[hbt]
    \centering    \includegraphics[width=0.9\textwidth,keepaspectratio]{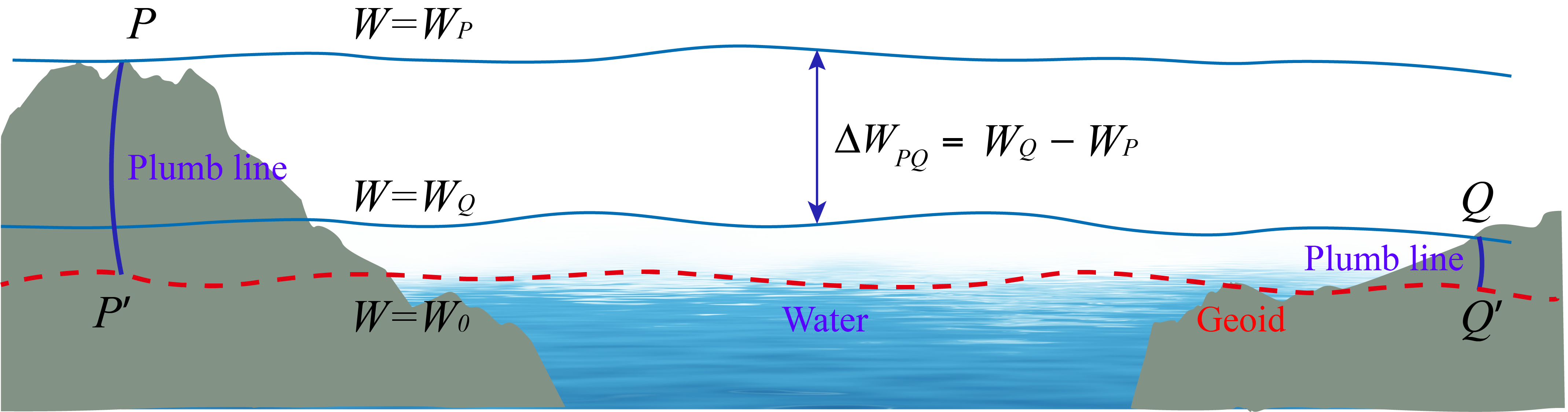}
    \caption{Red dashed red curve denotes the global geoid, the two solid blue curves denote the $W=W_P$ and $W=W_Q$ surfaces, respectively. Bold blue curve denotes the plumbline, along which the height integration is executed.}
    \label{fig:plumbline-int}
\end{figure}

The OH is a geometric length measured along the plumb-line from the ground point $i (i=P,Q)$ to its corespondent point $i'$  on the geoid ($i'=P', Q'$, see Fig. \ref{fig:plumbline-int}). 
For the OH case the $\hat g$ in Eq. \eqref{eq:H} is expressed as
\begin{equation}
    \hat g = \bar g = \frac{1}{H_P}\int_0^{H_P} g(h)dH_P ,
    \label{eq:hg}
\end{equation}
where $\bar g$ is the "mean value" of gravity $g(h)$ along the plumb-line. 
The normal height and dynamic height are approximations of OH \citep{Hofmann-Wellenhof2005-ur}, and OH is for practical purposes the height above sea level (in fact the height above the geoid). In this paper we will regard OH as the vertical coordinates of VHS.

Currently there are many different local VHSs which are difficult to be unified because the global mean sea level is not an equi-geopotential surface.
An important component of realization IHRS is to unify the global VHS, which requires defining a global reference surface that is assumed to be available all over the world \citep{Ihde2017-wk}.

\subsection{SFST method}
\label{ssec:SFST}
According to the general theory of relativity, we have the following relationship between the geopotentials $W_P$ and $W_Q$ and the clock frequencies $f_P$ and $f_Q$ for two points $P$ and $Q$ \citep{Weinberg1972-le,Bjerhammar1985-bd}
\begin{equation}
    \frac{f_P^2}{f_Q^2} = \frac{1 - 2W_Q/c^2}{1 - 2W_P/c^2} ,
\end{equation}
where $c$ is the speed of light in vacuum. Since the Earth's gravity field is weak, we have the following approximation
\begin{equation}
    W_P - W_Q = \frac{f_P - f_Q}{f} c^2 + O(c^{-4})
    \label{eq:bj} ,
\end{equation}
where $f = (f_P + f_Q)/2$, and $O(c^{-4})$ denote higher order terms which can be neglected if the stations $P$ and $Q$ are stationary near the Earth's surface.

Eq.\eqref{eq:bj} is sufficient for fibre link measurement between two clocks located on ground. 
However, when we use microwave links to compare two clocks located respectively on board a satellite and at a ground station, the case is much more complex.
For example, a satellite might move in high velocity which gives rise to big Doppler effects. 
In addition,
the ionosphere and troposphere will influence the frequency of microwaves propagating in space with medium, and the rotation and tidal effects of the Earth will change the status and environments of ground station.
In order to address these problems, recently we formulated the SFST approach for determining the geopotential difference between a satellite and a ground station or between two ground stations \citep{Shen2016-lc}. 
The main idea and formulations are briefly introduced as follows, details of which are referred to \citet{Shen2016-lc,Shen2017-kg}.

\begin{figure}[hbt]
  \centering
  \includegraphics[width=0.9\textwidth,keepaspectratio]{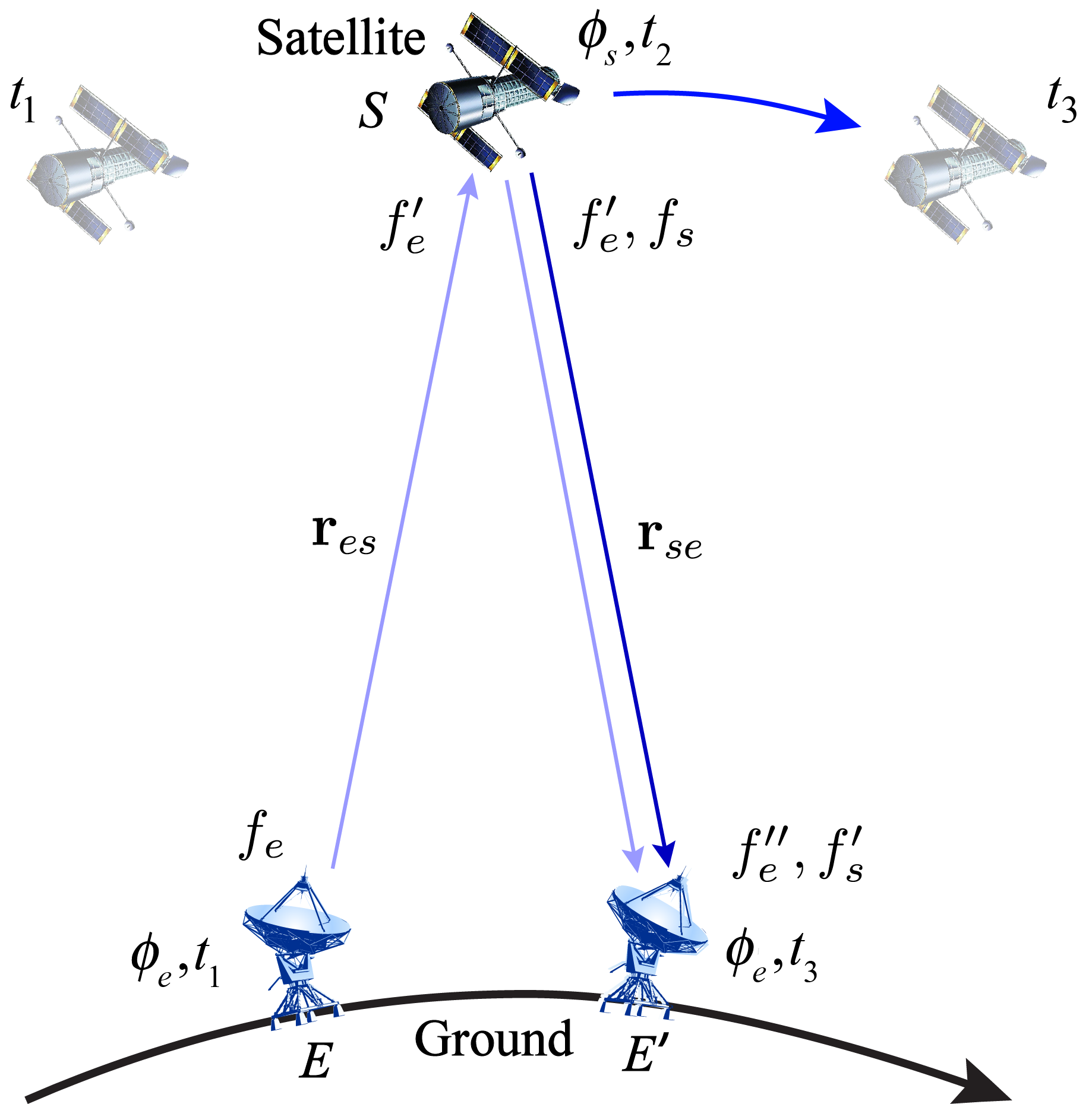}
  \caption{Ground station $E$ emits a frequency signal $f_e$ at time $t_1$, denoted by uplink (blue line). Satellite $S$ transmits the received signal $f_e'$ (the downlink denoted by blue line) and emits a new frequency signal $f_s$ at time $t_2$ (the downlink denoted by dark-blue line). The ground station receives signals $f_e''$ and $f_s'$ at time $t_3$ at position $E'$. $\phi$ is gravitational potential (GP), $\vec r$ is position vector.}
  \label{fig:orbit}
\end{figure}

Referring to Figure \ref{fig:orbit}, the SFST contains three microwave links. 
An emitter at a ground station $E$ emits a frequency signal $f_e$ at time $t_1$. 
When the signal is received by a satellite $S$ at time $t_2$, it immediately transmits the received signal $f_e'$ and emits a frequency signal $f_s$ simultaneously. 
These two signals that are simultaneously transmitted and emitted from the satellite are received by a receiver at ground station $E$ at time $t_3$, which are noted as $f_e''$ and $f_s'$, respectively. 
During the period of the emitting and receiving, the position of the ground station in space has been changed from $E$ to $E'$. The satellite transmits and emits signals at the same instant as it receives signal, so its position in the signal links is supposed to be the point S at time $t_2$. There might be a small amount of latency, due to the fact that during the transmitting, the position of the satellite is slightly different at the time as it receives and emits the signals. However, the un-synchronization influence is very small, which can be neglected for the SFST as explained in \cite{Shen2017-kg}).
The output frequency shift $\Delta f$ is expressed by a combination of three frequencies as \citep{Vessot1979-ot,Vessot1980-ax}
\begin{equation}
  \frac{\Delta f}{f_e}  = \frac{f_s'-f_s}{f_e}-\frac{(f_e''-f_e')+(f_e'-f_e)}{2f_e} ,
  \label{eq:frq}
\end{equation}

The beat frequency $\Delta f$ as expressed by Eq.\eqref{eq:frq} has cancelled out the first-order Doppler effect due to the relative motion between satellite and ground station.
As for the second-order Doppler effect and Earth's rotation influence, it is expressed as \citep{Vessot1979-ot}
  \begin{equation}
    \frac{\Delta f}{f_e} = \frac{\phi _s - \phi _e}{c^2} - \frac{\left| \vec v_e - \vec v_s \right| ^2}{2c^2} - \frac{\vec r_{se}\cdot \vec a_{e}}{c^2} + O(c^{-3}) ,
    \label{eq:ves}
  \end{equation}
where $\phi _s - \phi _e$ is the gravitational potential difference between the satellite and the ground station, $\vec v_e$ and $\vec v_s$ are velocities of ground station and satellite (spacecraft) respectively, $\vec r_{se}$ is vector from satellite to ground station, $\vec a_e$ is centrifugal acceleration vector of ground station, and $O(c^{-3})$ denote higher order terms than  $c^{-2}$.
On the right hand side of Eq.\eqref{eq:ves}, the second term denotes the second-order Doppler shift predicted by special relativity, and the third term represents the effect of Earth’s rotation during the signal's propagation time.

If the {higher order terms $O(c^{-3})$ are} omitted, Eq.\eqref{eq:ves} holds only at the accuracy level of $10^{-15}$ \citep{Cacciapuoti2011-oa}, and in this case it need not consider other influence factors such as the residual ionospheric effects, tidal effect etc.
To achieve one-centimeter level measurement in height, we considered higher order terms until $O(c^{-4})$ and various influence factors for satellite-ground microwave links, and derived a theoretical formula that holds at the accuracy level better than $10^{-18}$, expressed as \citep{Shen2017-kg}
\begin{equation}
    \frac{\Delta \phi_{es}}{c^2} \equiv  \frac{\phi_s-\phi_{e}}{c^2} = \frac{\Delta f}{f_e} - \frac{v_s^2 - v_{e}^2}{2c^2} - \sum^4_{i=1} q^{(i)} + \Lambda f + \delta f + O(c^{-5}) ,
    \label{eq:main}
\end{equation}
where $\Lambda f$ is the sum of all correction terms (it contains corrections of ionospheric and tropospheric effects, tidal effects and influence of celestial bodies), $\delta f$ is the sum of all error terms, $q^{(i)}$ ($i=1,2,3,4$) are quantities related to the positions and velocities of the ground station and satellite, second Newtonian potential, vector potential, and higher-order post Newtonian terms. The order terms higher than $O(c^{-5})$  are safely omitted.
The detailed expressions of the relevant quantities are referred to \citet{Shen2017-kg}. 
Based on Eq.\eqref{eq:main}, when the output frequency shift $\Delta f$ is measured and relevant quantities (such as position, speed, and acceleration of ground station and satellite) are given, the gravitational potential difference $\phi_{es}$ can be obtained.
We also discussed how to determine the geopotential difference between two ground stations, which are connected to a same satellite simultaneously \citep{Shen2017-kg}.
Since the satellite can serve as a "bridge" to connect the two ground sites, the geopotential difference between these two sites can be obtained.
Simulation experiments show that the precision of the geopotential difference between two ground sites determined by SFST method is about $2\sim 5$ cm in height \citep{Shen2017-kg}, under the assumption that the accuracy of OACs is $1\times 10^{-18}$,  which has been achieved recently \citep{McGrew2018-og,Huang2019-ez,Oelker2019-wm}.

\subsection{Determination of height difference between two ground height datum stations}
\label{subsec:2.3}
\begin{figure}
    \centering
    \includegraphics[width=0.9\textwidth,keepaspectratio]{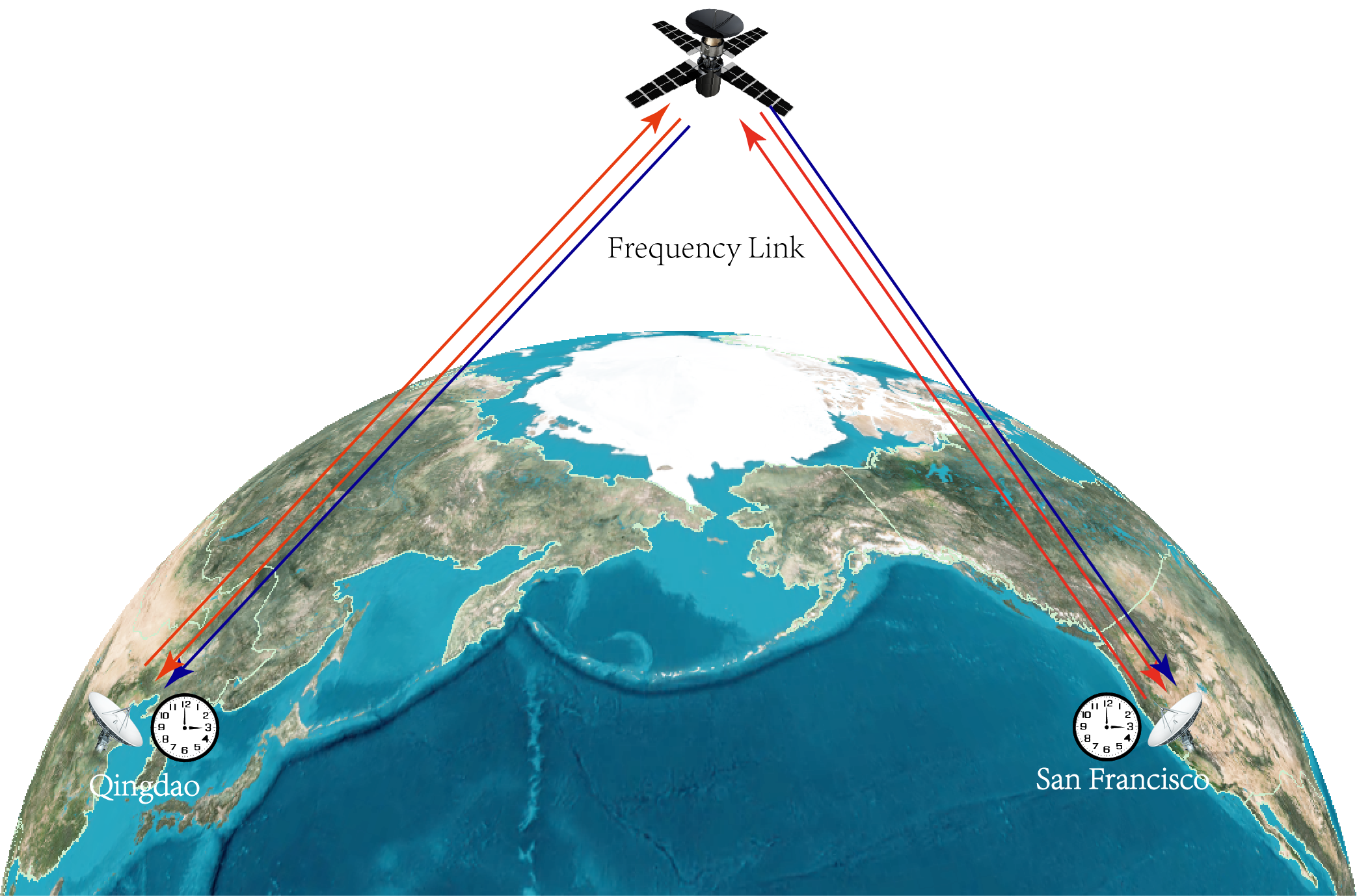}
    \caption{Connection of China HS originated at Qingdao datum and USA HS originated at San Francisco datum via satellite frequency signal transmission.}
    \label{fig:Satellite-Huni}
\end{figure}

Suppose we have two ground datum stations, Chinese height datum station at Qingdao and American height datum station at San Francisco (which were assumed), denoted respectively as $P$ and $Q$, located on two continents which are connected to the same satellite via SFST links simultaneously, cf. Fig. \ref{fig:Satellite-Huni}.

Suppose the gravitational potential difference $\Delta \phi_{PQ}$ between the datum points in China and  US has been determined using the SFST approach as described in section \ref{ssec:SFST}, then we may determine the geopotential difference $\Delta W_{PQ}$ by the following equation 
\begin{equation}
    \label{eq:geop}
    \Delta W_{PQ} = (\phi_{Q} - \phi_{P}) + (Z_{Q} - Z_{P}) ,
\end{equation}
where  $Z_{P}$ and $Z_{Q}$ are centrifugal force potentials at $P$ and $Q$, respectively; and $Z$ is expressed as 
\begin{equation}
    \label{eq:Qeq}
    Z = \frac{1}{2}\omega^2 (x^2+y^2) ,
\end{equation}
where $\omega$ is the angular velocity of the Earth rotation, $x$ and $y$ are coordinates defined in the geocentric Earth-fixed Cartesian coordinate system $o-xyz$ (e.g. ITRF2008, see \citet{Petit2010-hu}).



Suppose the height of point $P$ (noted as $H_P$) is given, and the geopotential difference $\Delta W_{PQ}$ has been measured by SFST method; then the height of point $Q$ (noted as $H_Q$) can be determined based on Eqs. \eqref{eq:H} and \eqref{eq:hg}, expressed as  
\begin{equation}
    \begin{split}
        H_Q &= \frac{W_0 - W_Q}{\bar g_Q} = \frac{W_0 - W_P - \Delta W_{PQ}}{\bar g_Q} , \\
        H_P &= \frac{W_0 - W_P}{\bar g_P} ,
    \end{split}
    \label{eq:height-det}
\end{equation}
where $\bar g_P$ and $\bar g_Q$ are the average gravity values along the plumb-lines $PP'$ and $QQ'$, respectively (see Fig. \ref{fig:plumbline-int}). 
It should be noted that $\bar g_i (i=P, Q)$ can not be directly calculated by Eq. \eqref{eq:hg}, because we do not know exactly the density distribution as well as the gravity distribution $g(h)$ inside the Earth.



We can see that besides the influence of the given value of $H_P$, the accuracy of the determined $H_Q$ depends on that of $\Delta W_{PQ}$; consequently it is related to the stabilities of the optical atomic clocks.
Since we cannot precisely determine the ``mean value'' $\bar{g}(i)$, in practical applications in plain region, $\bar{g}_i$ is usually replaced by the following formula \citep{Hofmann-Wellenhof2005-ur}
\begin{equation}
    \label{eq:g-bar}
    \bar{g}_i = g_i + 4.24\times 10^{-5} H_i ,
\end{equation}
where $g_i$, in gals (cm/s$^2$), is the gravity at ground point $i$, which can be measured by absolute gravimeter, and $H_i$, in meters, is the height difference between $i$ and $i' (i'=P', Q')$  (see Fig. \ref{fig:plumbline-int}). 
Therefore according to Eqs. \eqref{eq:height-det} and \eqref{eq:g-bar}, we obtain a practical formula for determining $H_Q$, expressed as
\begin{equation}
    \label{eq:H-prac}
    H_Q = \frac{H_P\cdot (g_P + 4.24\times 10^{-5} H_P) - \Delta W_{PQ}}{g_Q + 4.24\times 10^{-5} H_Q} ,
\end{equation}
where $\Delta W_{PQ}$ applies the geopotential unit (g.p.u, 1 g.p.u=1,000 gal.m), and iteration procedure could be applied if needed.

For the purpose of connecting VHSs, since the heights $H_P$ and $H_Q$ of the two height datum stations are relatively small (say less than 100 m), using Eq.\eqref{eq:H-prac} is sufficient. 
For instance, suppose $H_P=0$, $\Delta W_{PQ}=-100,000$ gal.m (which is equivalent to 100 m near Earth's surface), the maximum error caused by using Eq.\eqref{eq:H-prac} will not exceed $ 0.4$ mm.  
The reason is stated as follows. In the mentioned case,  $|H_Q| = \Delta W_{PQ}/(g_Q + 0.00424)$. The error caused by the uncertainty $\delta \bar{g}_i $ of the chosen mean gravity $\bar{g}_i $ will not exceed $0.00424$ gal. Then, we have $|\delta H_{Q}|=(\Delta W_{PQ}/g_{Q}^{2}) \delta \bar{g}_i =100 m \delta \bar{g}_i /g_{Q}\le  100 m \times 0.00424 gal/1000 gal=0.4$ mm. %

Now the height of the site $Q$ (in US) is determined under the same basis (geoid) as is that of the site $P$ (in China). 
Therefore, these two local VHSs are unified.
Based on the same principle, the SFST method can also be applied in the establishment of regional height system, and the geopotential difference (or height difference) of arbitrary two points located in this region can be directly determined, solving the regional height system (regional geoid) tilt problem.


\section{Simulation Experiments}
\label{sec:SE}
\begin{figure}
    \centering
    \includegraphics[width=0.9\textwidth,keepaspectratio]{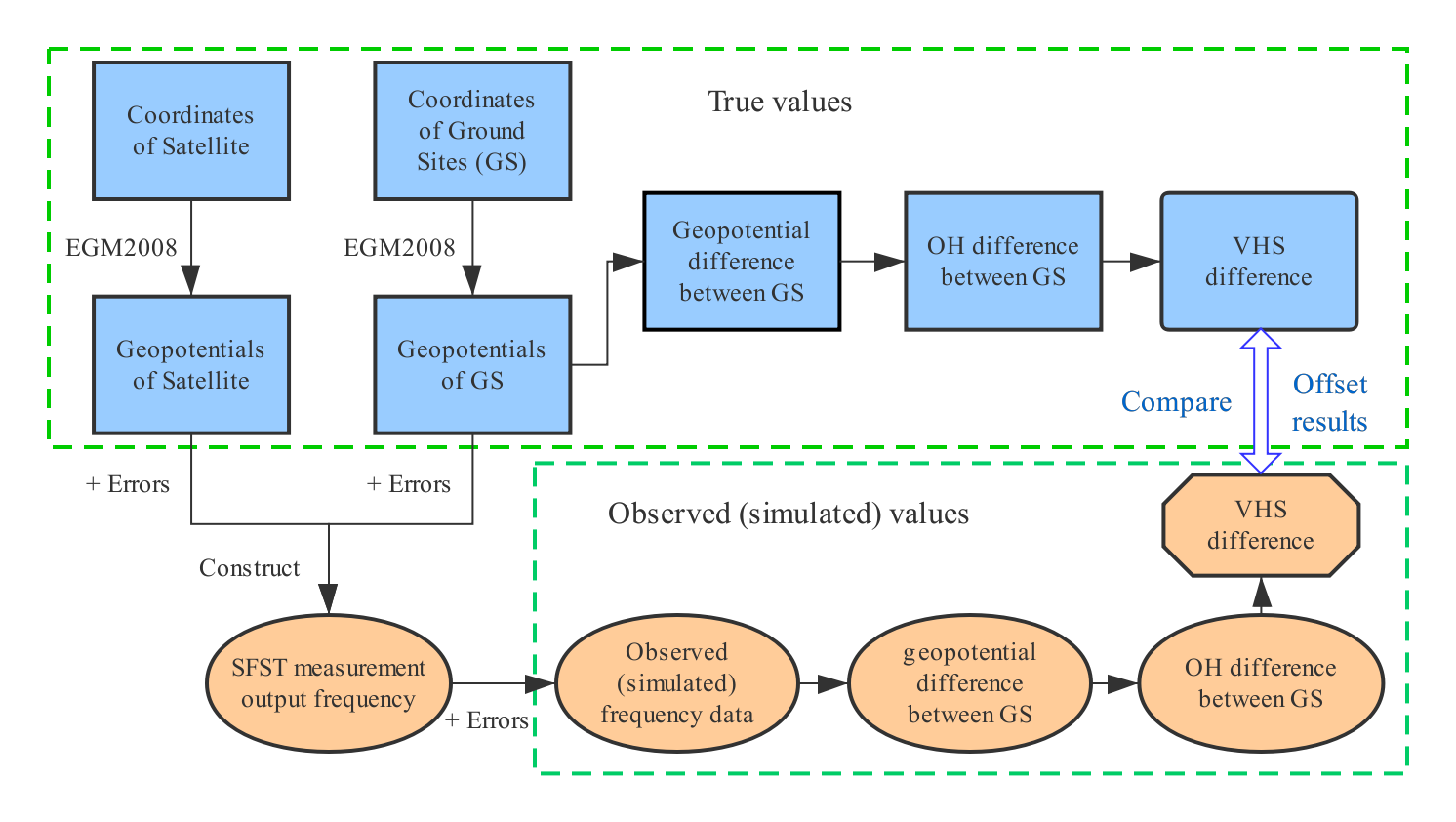}
    \caption{The scheme of the simulation experiment.}
    \label{fig:scheme}
\end{figure}

In this section we conducted simulation experiments to verify the SFST method of connecting two VHSs.
The main idea of the experiment is to compare a set of true values to a set of simulated observation values, as depicted in Fig. \ref{fig:scheme}, which is explained in the following subsections. 

\subsection{Input data}
\begin{figure}
    \centering
    \includegraphics[width=0.9\textwidth,keepaspectratio]{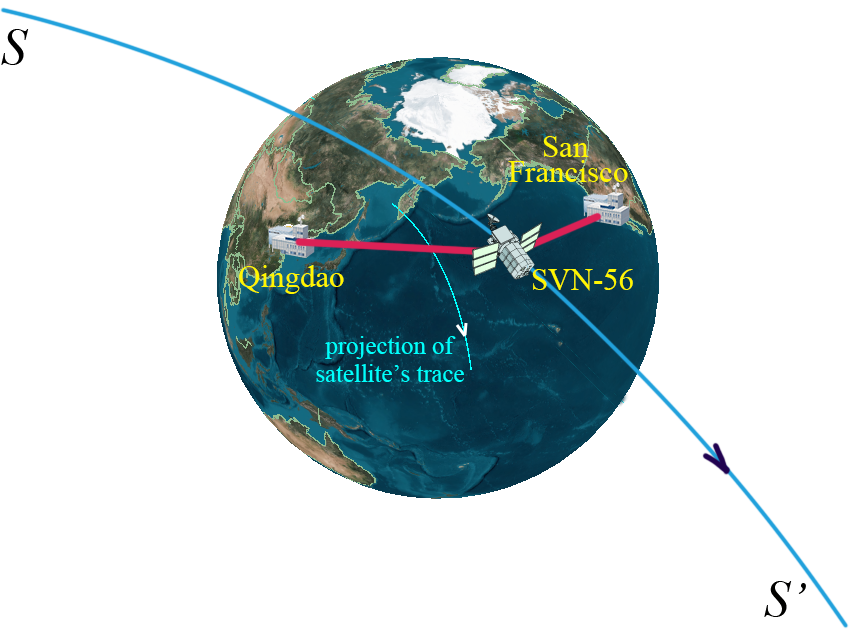}
    \caption{Experiments are conducted at the time duration  when the satellite SVN-56 moves from position $S$ to position $S'$ (from 7:00 am to 8:30 am, March 30, 2019).}
    \label{fig:exp}
\end{figure}

We chose two datum stations, Qingdao Datum Station (QDDS, which is located at Qingdao Guanxiangshan mountain and served as a height reference datum of China's VHS) and San Francisco Datum Station (SFDS,  which is located at California Academy of Science and supposed to be the height datum station of US's VHS), and connected them via a GNSS-type satellite, referring to Fig. \ref{fig:exp}.
The experiment time span is set for 1.5h, from 7:00 am to 8:30 am, March 30, 2019.
The satellite should be inter-visible to both the two ground datum stations during the experiment time; thus we chose the GPS navigation satellite SVN-56 which satisfies the requirement. The trace of SVN-56 during our experiment time is depicted in Fig. \ref{fig:exp} 

The orbit information of the GPS navigation satellite SVN-56 was obtained from the precise ephemeris provided by IGS(\url{www.igs.org/products}), which is regarded as true values.
The given coordinates of QDDS and SFDS (which are also regarded as true values) can be transferred from LLA to ECEF positions for later calculations.
The frequency links are designed to be established for every 5 second, hence we get a set of observation values for every 5 second.
Since the time interval between two data set in the precise ephemeris is 15 min, we use polynomial interpolation \citep{Horemuz2006-yr} to acquire the data set in 5-second interval (true values).
Then we use EGM2008 model \citep{Pavlis2012-fw} to calculate the gravitational potential values of satellite and two ground sites corresponding to the ``observation'' time points. 
These gravitational potential values are {regarded} as true values, namely the errors caused by EGM2008 are not considered (the accuracy of EGM2008 is about 10-20 cm at ground, and better than 1 cm at GNSS satellite altitude).
Then the true value of the geopotential difference between QDDS and SFDS, $\Delta W_{QD-SF}$, can be obtained. 

The microwave signals' frequencies will be affected by ionosphere and troposphere medium. 
Hence we use the International Reference Ionosphere Model \citep{Rawer1978-fl,Bilitza2017-zr} to obtain the electron density values, and Earth Global Reference Atmospheric Model \citep{Leslie2011-by} to obtain the temperature and pressure values, which are used to estimate the ionospheric and tropospheric influences on the signals' frequencies \citep{Namazov1975-qa,Millman1984-ni}.
The height and geopotential of the two ground sites will be also influenced by periodical tidal effect, which is well modeled \citep{Voigt2017-qc} and can be removed by some mature softwares such as ETERNA \citep{Wenzel1996-tk} and Tsoft \citep{Van_Camp2005-qt}.
In our experiment we use ETERNA to generate and analyses tide signals. We also considered the influences of other planets (such as Venus, Jupiter etc.) besides the Sun and the Moon. The relevant planet correction models are referred to \cite{Shen2017-kg}.

In our experiment, the two datum stations are connected to SVN-56 simultaneously via SFST links.
Relevant input parameters are listed in Table \ref{tab:pre}.
It should be noted that the OH of QDDS is released by Chinese government as China's height datum origin, but the US has no corresponding height datum origin. 
Therefore the OH of SFDS is deduced from EGM2008. 
We assume the height difference between China's VHS and US's VHS is $1.000$ m (as assumed true value), and China's VHS is higher than US's VHS.

\begin{table}
    \caption{The input datas used in simulation experiments. The coordinates are based on ITRF14}
    \label{tab:pre}
    \footnotesize
    \begin{tabular}{@{}lll}
        \hline\noalign{\smallskip}
        Entities & & Values of Parameters \\
        \noalign{\smallskip}\hline\noalign{\smallskip}
        Satellite & ID & SVN-56 (GPS Navigation Sat.) \\
        & Coord. & from (-19167.235509, 3652.729794, 18038.749481) \\
        & &to (-26493.102586, 424.868409, 3830.004962) \\
        Qingdao DS & LLA & (36.06974$^\circ$N, 120.32172$^\circ$E, 77.472 m) \\ 
        & ECEF (m) & (-2605813.108, 4455436.499, 3734494.956) \\ 
        & OH (m) & 72.260 \\
        San Francisco DS & LLA & (37.76985$^\circ$N, 122.46616$^\circ$W, 75.878 m) \\
        & ECEF(m) & (-2709867.959, -4259189.792, 3885328.909) \\
        & OH (m) & 109.126 \\ 
        Gravity field model & & EGM2008 \\
        Ionospheric model & & International Reference Ionosphere \\
        Tropospheric model & & Earth  Global  Reference Atmospheric Model \\
        Tide correction & & ETERNA \\
        Observation duration & & from 7:00 am to 8:30 am, March 30, 2019 \\
        Mearsurement interval & & 5 s \\
        Height systems diff. & & 1.000 m (China HS is higher than US HS)\\
    \noalign{\smallskip}\hline\noalign{\smallskip}
    \end{tabular}
\end{table}

\subsection{``Observed'' values}
According to Eqs. \eqref{eq:main} and \eqref{eq:geop}, the geopotential difference between QDDS and {SFDS, $\Delta \hat W_{QD-SF}(t) $,} can be measured as time series
\begin{equation}
    \Delta \hat W_{QD-SF}(t) = \Delta \hat \phi_{QD-s}(t) - \Delta \hat \phi_{SF-s}(t) + (Z_{SF} - Z_{QD}) ,
    \label{eq:QD-SF}
\end{equation}
where $\Delta \hat \phi_{QD-s}(t)$ {and  $\Delta \hat \phi_{SF-s}(t)$  are respectively} the observed gravitational potential {differences between QDDS and the satellite as well as SFDS and the satellite, at time $t$,} $Z_{QD}$ and $Z_{SF}$ are centrifugal force potentials of QDDS and SFDS respectively.

The observed values $\Delta \hat W_{QD-SF}(t)$ are different from true geopotential difference value $\Delta W_{QD-SF}$ because they are influenced by various error sources.
In this simulation experiment we have considered clock error $e_{clk}$, ionosphere residual error $e_{ion}$, troposphere residual error $e_{tro}$, satellite's position and velocity errors $e_{pos}$ and $e_{vel}$, tidal correction residual error $e_{tide}$, and asynchronism error $e_{asy}$. We expect that $\Delta \hat \phi_{QD-s}(t)$ and $\Delta \hat \phi_{SF-s}(t)$ are measured at the same time $t$, but in practice they might have slight difference, which will introduce the asynchronism error.
    The above mentioned various errors are considered as noises, which are added to the true values. The total errors $e_{all}$ are expressed in the following form
\begin{equation}
    e_{all} = e_{clk} + e_{ion} + e_{tro} + e_{pos} + e_{vel} + e_{tide} + e_{asy} ,
    \label{eq:e-all}
\end{equation}
 The magnitude and behavior of each kind of error play important role in this experiment; thereby we need to investigate different error models based on different error sources to make the simulation case more close to the real case.

Currently the most precise OACs have demonstrated $4.8\times 10^{−17}$ stability at 1 second, and $6.6\times 10^{−19}$ in 1 hour for two clocks comparison \citep{Oelker2019-wm}; therefore we set the error magnitude of $e_{clk}$ as $4.8\times 10^{−17}$.
Although there are many kinds of random noises that affect OAC signals \citep{Major2013-qc}, the most prominent components are white frequency modulation and random walk frequency modulation \citep{Galleani2003-hu}.
Correspondingly the behaviors of clock errors are modeled as the following equation
\begin{equation}
    e_{clk}(t) = a_{clk} + b_{clk}\cdot t + c_{clk}\cdot \phi (t) + d_{clk}\cdot \int_0^t \xi (t)dt ,
    \label{eq:e-clk}
\end{equation}
where $a_{clk}$, $b_{clk}$, $c_{clk}$ and $d_{clk}$ are constant coefficients, $\phi (t)$ and $\xi (t)$ are both standard white Gaussian noises.
Each term in the right side of Eq.\eqref{eq:e-clk} has clear physical meaning; specifically $a_{clk}$ denotes the initial frequency difference, $b_{clk}\cdot t$ is the drift term, $c_{clk}\cdot \phi (t)$ is the white noise component, and $d_{clk}\cdot \int_0^t \xi (t)dt$ represents the random walk effect.
As we set proper values of constant coefficients in accordance with the performance of OACs in \citet{Oelker2019-wm}, a series of frequency comparison data with errors embedded can be generated.

For other error sources, their magnitudes are discussed in detail in \citet{Shen2017-kg} and listed in Table \ref{tab:err}.
It should be noted that though the residual influences of ionosphere and troposphere for SFST method are at the centimeter level (corresponding to frequency shift of $10^{-18}$ level), we have established correction models \citep{Shen2017-kg} to reduce their influences to the millimeter level.
The Earth tide effects could reach up to 60 cm at maximum \citep{Poutanen1996-nx}, but the residual error in vertical direction after corrections can be limited to 2 mm for solid Earth tide \citep{Li2018-mh}, and 8 mm for ocean tide loading \citep{Penna2008-ji}.

Since there are no mature mathematical models for above mentioned errors and their influences are much smaller than the clock errors (see Table. \ref{tab:err}), we adopt a general error model which contains systematic (initial) offset, drift and white Gaussian noises for each of the error source, expressed as the following equation
\begin{equation}
    e_j(t) = a_j + b_j\cdot t + c_j\cdot \phi_i (t) ,~~~~~(j=ion, tro, pos, vel, tide, asy)
    \label{eq:e-other}
\end{equation}
where $a_j$, $b_j$ and $c_j$ are constant coefficients, which are randomly set in accordance with the error magnitudes listed in {Table} \ref{tab:err}

\begin{table}
    \caption{Error magnitudes of different error sources in determining the gravitational potential difference between a satellite and a ground station. They are transformed to relative frequency (modified after \cite{Shen2017-kg})}
    \footnotesize
    \begin{tabular}{@{}lll}
      \hline\noalign{\smallskip}
    Influence factor & (Residual) Error magnitude in $\Delta f / f_e$ \\
    \noalign{\smallskip}\hline\noalign{\smallskip}
    ionospheric correction residual & $\delta f_{ion} \sim 5.5\times 10^{-19}$  \\
    tropospheric correction residual & $\delta f_{tro} \sim 1.9\times 10^{-19}$  \\
    tidal correction residual & $\delta f_{tide} \le 10^{-18}$ \\
    position \& velocity & $\delta f_{vepo} \sim 3.4\times 10^{-19}$ (10 mm and 0.1mm/s $^b$) \\
    asynchronism & $\delta f_{delay} \sim {10^{-19}}^c$ (below 1 ms) \\
    clock error & $\delta f_{osc} \sim 4.8\times {10^{-17}}$ \\
    \noalign{\smallskip}\hline\noalign{\smallskip}
    \end{tabular}
    \label{tab:err}
    \\
  $^a$ After tri-frequency combination; \\
  $^b$ Satellite's position errors are assumed as 10 mm \citep{Kang2006-ix}, velocity errors are assumed as 0.1mm/s \citep{Sharifi2013-bm}. \\
\end{table}

According to Eqs. \eqref{eq:e-clk} and \eqref{eq:e-other}, we can generate the noise signals based on the magnitudes and nature of the error sources at any time.
Then these noises are added to relevant true values, and we get a set of relevant "Observed" values, based on which the geopotential difference $\Delta \hat W_{QD-SF}(t)$ is determined using Eqs. \eqref{eq:main} and \eqref{eq:QD-SF}.
The next step is converting the geopotential difference to corresponding height difference.
Without loss of generality, assuming the zero-height surface of China's VHS is just coinciding with the $W_0$ surface, based on China's VHS, the height of SFDS can be calculated by Eq. \eqref{eq:H-prac}, expressed as
\begin{equation}
    \hat H_{SF}(t) = \frac{H_{QD}\cdot (g_{QD} + 0.0424H_{QD}) - \Delta \hat W_{QD-SF}(t)}{g_{SF} + 0.0424\hat H_{SF}(t)} ,
    \label{eq:Hst}
\end{equation}
where $H_{Q} = 72.260$ m is the height of QDDS in China's VHS.
In this case, the observed VHS difference between China and US can be obtained as
\begin{equation}
    \Delta \hat H_{VHS}(t) = \hat H_{SF}(t) - H_{SF} ,
    \label{eq:Hvhs}
\end{equation}
where $H_{SF} = 109.126$ m is the height of SFDS in US's VHS, and the unification of the two VHSs is realized. 
However, if the zero-height surface of China's VHS does not coincide with the $W_0$ surface (this is in general the real case), Eq. \eqref{eq:Hst} is not rigorous, and in this case Eq. \eqref{eq:height-det} should be modified as
\begin{equation}
    \begin{split}
    H_{QD}  &= \frac{W_0 + \delta W_{China} - W_{QD}}{\bar g_{QD}} , \\
    H_{SF}  &= \frac{W_0 + \delta W_{US} - W_{SF}}{\bar g_{SF}} , \\
    \hat H_{SF} &= \frac{W_0 + \delta W_{China} - W_{QD} - \Delta W_{QD-SF}}{\bar g_{SF}} ,
    \end{split}
    \label{eq:Hd-real}
\end{equation}
where $\delta W_{China}$ is the geopotential difference between the zero-height surface of China's VHS and the $W_0$ surface, $\delta W_{USA}$ is the geopotential difference between the zero-height surface of US's VHS and the $W_0$ surface.
If $\delta W_{China}$ is unknown, the derived height of SFDS $\hat H_{SF}(t)$ can not be calculated based on the height of QDDS $H_{QD}$ even though their geopotential difference is given; therefore the height difference between the two VHSs cannot be strictly determined.
However, as the $W_0$ surface and a VHS's zero-height surface are close to the mean sea level, $\delta W_{i} (i = China, US)$ are relatively small (usually less than 10 $\rm{m}^2/\rm{s}^{-2}$ \citep{Sideris2015-fn}); the error introduced by Eqs. \eqref{eq:Hst} and \eqref{eq:Hvhs} can be neglected at current precision level of centimeter.
In addition, if the value of $\delta W_i$ can be obtained (which is very promising in future, see discussions in Sec. \ref{sec:con}), we can also unify the two VHSs based on rigorous equation.
Therefore for brevity and without loss of generality, we just use Eqs. \eqref{eq:Hst} and \eqref{eq:Hvhs} for the height unification calculation.
By comparing the observed height difference $\Delta \hat H_{VHS}(t)$ and the true difference $\Delta H_{VHS} = 1$ m, the reliability of SFST approach for height system unification can be verified.

\subsection{Experiment results}
\begin{figure}
    \centering
    \includegraphics[width=0.9\textwidth,keepaspectratio]{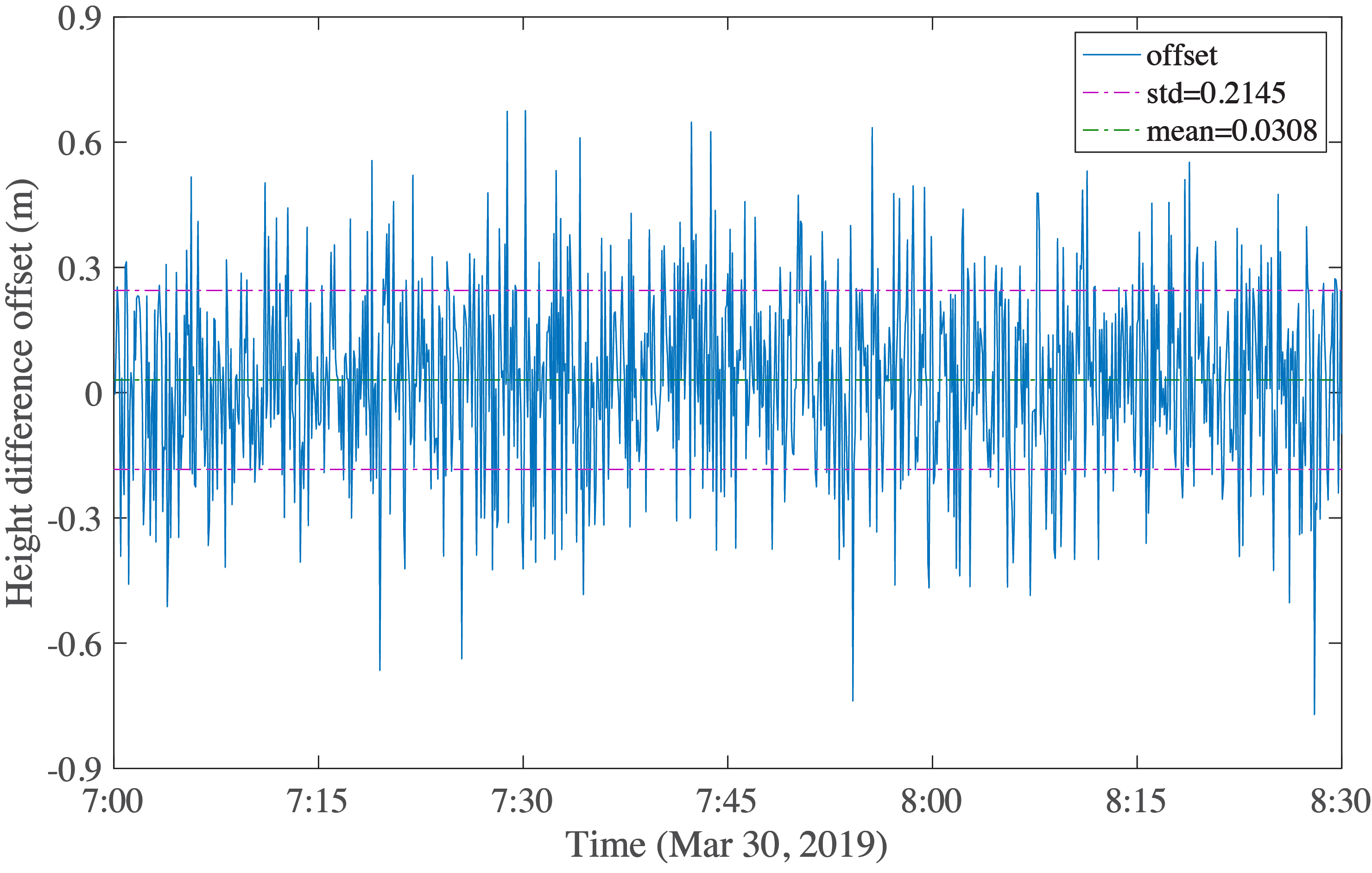}
    \caption{The offset between true values and estimated values of Height datum difference determined by SVN-56 satellite.}
    \label{fig:trend}
\end{figure}

Since the experiment time span {in one times (from 7:00 to 8:30, on March 30, 2019)}   is {1.5 h} and the measurement interval is 5 s, there are 1080 observation values in total.
The results {in first experiment (namely from 7:00 to 8:30, on March 30, 2019)} are shown in Fig. \ref{fig:trend}, with the mean offset 3.08 cm and the STD value of 21.45 cm {(see the first row of Table \ref{tab:res}).}

We can see that although the STD is relatively large, at the decimeter level, the mean offset value is small, at the centimeter level.
That is because the main component of the clock errors is white noises, while the drift and random walk effects are not quite obvious in the results.
Therefore, since the stability of the clock can be significantly improved after a period (say one hour) of integration, as demonstrated by \cite{Oelker2019-wm},  the height unification accuracy could be improved after a period of integration.

\begin{figure}
    \centering
    \includegraphics[width=0.9\textwidth,keepaspectratio]{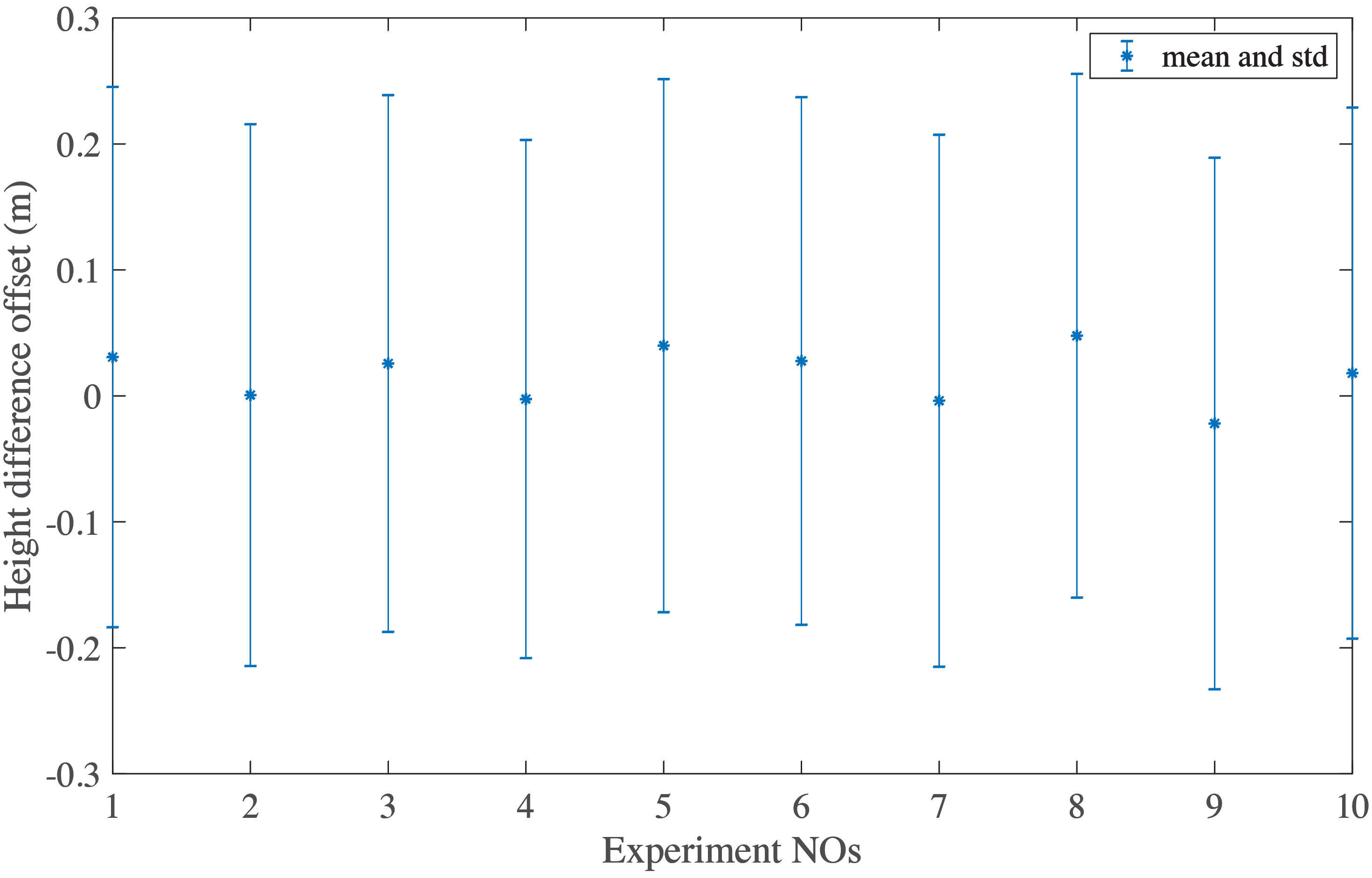}
    \caption{The mean offset values and STD values of Height datum difference in 10 different simulation experiments.}
    \label{fig:10}
\end{figure}

In order to improve the accuracy of the results, we may use multi-observations, namely we may use observations in different time periods in one day or different days.
To improve the precision, with different randomly chosen coefficients $a_i$, $b_i$, $c_i$ and $d_i$ in Eqs. \eqref{eq:e-clk} and \eqref{eq:e-other} we run 10 times simulation experiments in total. The behaviors of the offset signals are similar. Thus we only display the mean offset values and STD values in Fig. \ref{fig:10}.
We can see that the mean offsets are limited to centimeter level, and the largest mean offset is 4.77 cm in the 8th experiment.
The final results are listed in Table \ref{tab:res}.

\begin{table}
    \caption{The results of 10 simulation experiments. Relevant parameters are listed in Table \ref{tab:pre}}
    \label{tab:res}
    \footnotesize
    \begin{tabular}{@{}llll}
        \hline\noalign{\smallskip}
        Experiment & Height diff. between China's  & Offset to true & STD \\
        No. & VHS and US' VHS (m)  &  value (1 m)  & (m) \\
        \noalign{\smallskip}\hline\noalign{\smallskip}
        1 & 1.0308 & 0.0308 & 0.2145 \\
        2 & 1.0061 & 0.0061 & 0.2151 \\
        3 & 1.0257 & 0.0257 & 0.2130 \\
        4 & 0.9975 & -0.0025 & 0.2056 \\
        5 & 1.0399 & 0.0399 & 0.2116 \\
        6 & 1.0277 & 0.0277 & 0.2094 \\
        7 & 0.9961 & -0.0039 & 0.2112 \\
        8 & 1.0477 & 0.0477 & 0.2079 \\
        9 & 0.9781 & -0.0219 & 0.2110 \\
        10 & 1.0180 & 0.0180 & 0.2108 \\
        \hline\noalign{\smallskip}
        Average & 1.0168 & 0.0168 & 0.2110\\
    	\noalign{\smallskip}\hline\noalign{\smallskip}
    \end{tabular}
\end{table}

In practical applications, we may use several days' data to estimate the height difference. For instance, if 10 different experiments are conducted in different time periods (e.g. continuous 10 days, every day from 7:00 to 8:30), the final results can be improved after taking average.

\section{Conclusions}
\label{sec:con}
In this study we formulated an approach to unify different local vertical height systems in centimeter level via ultra-high precise frequency signal links between one satellite and two datum stations separated by oceans, and performed simulation experiments addressing this issue by taking an example of connecting the China's VHS and the US's VHS based on the SFST approach. 
Simulation experiment results show that,  the deviation between the calculated result based on the ``observations''  and the true result is around 2 cm, with an accuracy level (STD) of 2 decimeters in 1.5 h, provided that the OACs' stability achieves the level of $4.8\times 10^{-17}$ in one second. 
Results of simulation experiments confirmed the reliability of the SFST approach in unifying VHSs, and the precision could be greatly improved by more observations in more time periods.

A prerequisite for the SFST approach is frequency synchronization before the measurement, which means that the output frequency of OACs' oscillators should be identical if their locations have the same geopotential value.
The error of a prior synchronization will also affect the precision of height unification based on relativistic geodetic methods.
Two clocks can be easily synchronized in the same position; but when they are separated at large distance as in our case, it is very challenging to precisely synchronize them.
However, it can be realized by a combined method of fiber connection and repeated clock transportation.
How to precisely synchronize two separated clocks is a meticulous technique problem, which is not quite relevant to the main topic of this study. Hence, we will address that problem in a separated paper.

With quick development of time-frequency science, ultra-high precise OACs (say at $1\times 10^{-18}$ level or better within one hour) have been developed and still under improvement, which make the SFST approach prospective for the unification of VHSs at centimeter level.
The SFST approach is also prospective for realizing the IHRS.
As a preliminary study we only connect two stations in this work. 
However, if a globally covered SFST network is established, the VHSs all over the world can be unified.

Compared to conventional methods, the main merits of SFST method lie in that it can directly determine the geopotential difference between two arbitrary stations in a relatively short  period (say several hours or one day), and it is not subjected to distance or obstacles such as mountains or oceans.
However it also has some limitations.
The first problem is the requirement of ultra-high precise clocks and relevant equipments, which makes the measurement relatively difficult currently.
Another problem is that its precision is slightly lower than the optical fibre links method.
Therefore at present, the best practice is to adopt the SFST method as a supplement of conventional methods. For example, we can use SFST method to connect the benchmarks of two VHSs far apart, and use conventional methods and optical fibre links method for local VHS unification.

\section*{Acknowledgements}
We would like to express our sincere thanks to J{\"u}ergen Kusche, and three anonymous Reviewers for their valuable comments and suggestions, which greatly improved the manuscript. 
This study is supported by NSFC (grant Nos. 41804012, 41631072, 41721003, 41574007 and 41429401), Natural Science Foundation of Hubei Province of China (grant No. 2019CFB611), the Discipline Innovative Engineering Plan of Modern Geodesy and Geodynamics (grant No. B17033), DAAD Thematic Network Project (grant No. 57173947) and ISSI 2017 Supporting Project.

\bibliography{Ref}

\begin{thebibliography}{}

\bibitem[Amjadiparvar et~al., 2016]{Amjadiparvar2016-lg}
Amjadiparvar, B., Rangelova, E., and Sideris, M.~G. (2016).
\newblock The {GBVP} approach for vertical datum unification: recent results in
  north america.
\newblock {\em Journal of Geodesy}, 90(1):45--63.

\bibitem[Bilitza et~al., 2017]{Bilitza2017-zr}
Bilitza, D., Altadill, D., Truhlik, V., Shubin, V., Galkin, I., Reinisch, B.,
  and Huang, X. (2017).
\newblock International reference ionosphere 2016: From ionospheric climate to
  real-time weather predictions.
\newblock {\em Space Weather}, 15(2):418--429.

\bibitem[Bjerhammar, 1985]{Bjerhammar1985-bd}
Bjerhammar, A. (1985).
\newblock On a relativistic geodesy.
\newblock {\em Bull. Am. Assoc. Hist. Nurs.}, 59(3):207--220.

\bibitem[Cacciapuoti and Salomon, 2011]{Cacciapuoti2011-oa}
Cacciapuoti, L. and Salomon, C. (2011).
\newblock Atomic clock ensemble in space ( {ACES} ).
\newblock {\em European Space Agency, (Special Publication) ESA SP},
  327(385):295--297.

\bibitem[Drewes et~al., 2016]{Drewes2016-lc}
Drewes, H., Kuglitsch, F., Ad{\'a}m, J., and R{\'o}zsa, S. (2016).
\newblock The geodesist's handbook 2016.
\newblock {\em J. Geodesy}, 90(10):907--1205.

\bibitem[Einstein, 1915]{Einstein1915-tv}
Einstein, A. (1915).
\newblock Die feldgleichungen der gravitation.
\newblock {\em Sitzungsberichte der K{\"o}niglich Preu{\ss}ischen Akademie der
  Wissenschaften (Berlin).}, 1:844--847.

\bibitem[Flury, 2016]{Flury2016-bw}
Flury, J. (2016).
\newblock Relativistic geodesy.
\newblock {\em J. Phys. Conf. Ser.}, 723(1):012051.

\bibitem[Galleani et~al., 2003]{Galleani2003-hu}
Galleani, L., Sacerdote, L., Tavella, P., and Zucca, C. (2003).
\newblock A mathematical model for the atomic clock error.
\newblock {\em Metrologia}, 40(3):S257.

\bibitem[Gerlach and Rummel, 2013]{Gerlach2013-tw}
Gerlach, C. and Rummel, R. (2013).
\newblock Global height system unification with {GOCE}: a simulation study on
  the indirect bias term in the {GBVP} approach.
\newblock {\em J Geod}, 87(1):57--67.

\bibitem[Grotti et~al., 2018]{Grotti2018-ap}
Grotti, J., Koller, S., Vogt, S., H{\"a}fner, S., Sterr, U., Lisdat, C.,
  Denker, H., Voigt, C., Timmen, L., Rolland, A., Baynes, F.~N., Margolis,
  H.~S., Zampaolo, M., Thoumany, P., Pizzocaro, M., Rauf, B., Bregolin, F.,
  Tampellini, A., Barbieri, P., Zucco, M., Costanzo, G.~A., Clivati, C., Levi,
  F., and Calonico, D. (2018).
\newblock Geodesy and metrology with a transportable optical clock.
\newblock {\em Nat. Phys.}, 14(5):437--441.

\bibitem[Hirt and Kuhn, 2012]{Hirt2012-vy}
Hirt, C. and Kuhn, M. (2012).
\newblock Evaluation of high-degree series expansions of the topographic
  potential to higher-order powers: {TOPOPOTENTIAL} {TO} {HIGHER-ORDER}
  {POWERS}.
\newblock {\em J. Geophys. Res.}, 117(B12).

\bibitem[Hofmann-Wellenhof and Moritz, 2005]{Hofmann-Wellenhof2005-ur}
Hofmann-Wellenhof, B. and Moritz, H. (2005).
\newblock {\em Physical geodesy}.
\newblock Springer.

\bibitem[Horemu{\v z} and Andersson, 2006]{Horemuz2006-yr}
Horemu{\v z}, M. and Andersson, J.~V. (2006).
\newblock Polynomial interpolation of {GPS} satellite coordinates.
\newblock {\em GPS Solutions}, 10(1):67--72.

\bibitem[Huang et~al., 2019]{Huang2019-ez}
Huang, Y., Guan, H., Zeng, M., Tang, L., and Gao, K. (2019).
\newblock {$^{40}$Ca$^{+}$} ion optical clock with micromotion-induced shifts
  below 10$^{-18}$.
\newblock {\em Phys. Rev. A}, 99(1):011401.

\bibitem[Ihde et~al., 2008]{Ihde2008-iu}
Ihde, J., M{\"a}kinen, J., and Sacher, M. (2008).
\newblock Conventions for the definition and realization of a european vertical
  reference system ({EVRS)--EVRS} conventions 2007.
\newblock {\em EVRS Conv}, pages 1--10.

\bibitem[Ihde et~al., 2017]{Ihde2017-wk}
Ihde, J., S{\'a}nchez, L., Barzaghi, R., Drewes, H., Foerste, C., Gruber, T.,
  Liebsch, G., Marti, U., Pail, R., and Sideris, M. (2017).
\newblock Definition and proposed realization of the international height
  reference system ({IHRS}).
\newblock {\em Surv. Geophys.}, 38(3):549--570.

\bibitem[Kang et~al., 2006]{Kang2006-ix}
Kang, Z., Tapley, B., Bettadpur, S., Ries, J., Nagel, P., and Pastor, R.
  (2006).
\newblock Precise orbit determination for the {GRACE} mission using only {GPS}
  data.
\newblock {\em J. Geodesy}, 80(6):322--331.

\bibitem[Kopeikin et~al., 2011]{Kopeikin2011-bo}
Kopeikin, S., Efroimsky, M., and Kaplan, G. (2011).
\newblock Relativistic geodesy.
\newblock In {\em Relativistic Celestial Mechanics of the Solar System},
  volume~83 of {\em Bulletin of the American Astronomical Society}, pages
  671--714. Wiley-VCH Verlag GmbH \& Co. KGaA, Weinheim, Germany.

\bibitem[Leslie and Justus, 2011]{Leslie2011-by}
Leslie, F.~W. and Justus, C.~G. (2011).
\newblock The {NASA} marshall space flight center earth global reference
  atmospheric model-2010 version.
\newblock Technical report.

\bibitem[Li et~al., 2018]{Li2018-mh}
Li, F., Lei, J., Zhang, S., Ma, C., Hao, W., E, D., and Zhang, Q. (2018).
\newblock The impact of solid earth-tide model error on tropospheric zenith
  delay estimates and {GPS} coordinate time series.
\newblock {\em Survey Review}, 50(361):355--363.

\bibitem[Lion et~al., 2017]{Lion2017-pl}
Lion, G., Panet, I., Wolf, P., Guerlin, C., Bize, S., and Delva, P. (2017).
\newblock Determination of a high spatial resolution geopotential model using
  atomic clock comparisons.
\newblock {\em J. Geodesy}, 91(6):597--611.

\bibitem[Lisdat et~al., 2016]{Lisdat2016-pz}
Lisdat, C., Grosche, G., Quintin, N., Shi, C., Raupach, S. M.~F., Grebing, C.,
  Nicolodi, D., Stefani, F., Al-Masoudi, A., D{\"o}rscher, S., H{\"a}fner, S.,
  Robyr, J.-L., Chiodo, N., Bilicki, S., Bookjans, E., Koczwara, A., Koke, S.,
  Kuhl, A., Wiotte, F., Meynadier, F., Camisard, E., Abgrall, M., Lours, M.,
  Legero, T., Schnatz, H., Sterr, U., Denker, H., Chardonnet, C., Le~Coq, Y.,
  Santarelli, G., Amy-Klein, A., Le~Targat, R., Lodewyck, J., Lopez, O., and
  Pottie, P.-E. (2016).
\newblock A clock network for geodesy and fundamental science.
\newblock {\em Nat. Commun.}, 7:12443.

\bibitem[Luz et~al., 2002]{Luz2002-nq}
Luz, R.~T., Fortes, L. P.~S., Hoyer, M., and Drewes, H. (2002).
\newblock The vertical reference frame for the americas --- the sirgas 2000
  {GPS} campaign ---.
\newblock In {\em Vertical Reference Systems}, pages 302--305. Springer Berlin
  Heidelberg.

\bibitem[Major, 2013]{Major2013-qc}
Major, F.~G. (2013).
\newblock {\em The Quantum Beat: The Physical Principles of Atomic Clocks}.
\newblock Springer Science \& Business Media.

\bibitem[McGrew et~al., 2018]{McGrew2018-og}
McGrew, W.~F., Zhang, X., Fasano, R.~J., Sch{\"a}ffer, S.~A., Beloy, K.,
  Nicolodi, D., Brown, R.~C., Hinkley, N., Milani, G., Schioppo, M., Yoon,
  T.~H., and Ludlow, A.~D. (2018).
\newblock Atomic clock performance enabling geodesy below the centimetre level.
\newblock {\em Nature}, 564(7734):87--90.

\bibitem[Mehlst{\"a}ubler et~al., 2018]{Mehlstaubler2018-da}
Mehlst{\"a}ubler, T.~E., Grosche, G., Lisdat, C., Schmidt, P.~O., and Denker,
  H. (2018).
\newblock Atomic clocks for geodesy.
\newblock {\em Rep. Prog. Phys.}, 81(6):064401.

\bibitem[Millman and Arabadjis, 1984]{Millman1984-ni}
Millman, G.~H. and Arabadjis, M.~C. (1984).
\newblock Tropospheric and ionospheric phase perturbations and doppler
  frequency shift effects.
\newblock {\em Nasa Sti/recon Technical Report N}, 85.

\bibitem[M{\"u}ller et~al., 2008]{Muller2008-pc}
M{\"u}ller, J., Soffel, M., and Klioner, S.~A. (2008).
\newblock Geodesy and relativity.
\newblock {\em J. Geodesy}, 82(3):133--145.

\bibitem[Namazov et~al., 1975]{Namazov1975-qa}
Namazov, S.~A., Novikov, V.~D., and Khmel'nitskii, I.~A. (1975).
\newblock Doppler frequency shift during ionospheric propagation of decameter
  radio waves (review).
\newblock {\em Radiophys. Quantum Electron.}, 18(4):345--364.

\bibitem[Oelker et~al., 2019]{Oelker2019-wm}
Oelker, E., Hutson, R.~B., Kennedy, C.~J., Sonderhouse, L., Bothwell, T.,
  Goban, A., Kedar, D., Sanner, C., Robinson, J.~M., Marti, G.~E., Matei,
  D.~G., Legero, T., Giunta, M., Holzwarth, R., Riehle, F., Sterr, U., and Ye,
  J. (2019).
\newblock Demonstration of 4.8 $\times$ 10$^{−17}$ stability at 1 s for two
  independent optical clocks.
\newblock {\em Nat. Photonics}, 13(10):714--719.

\bibitem[Pavlis et~al., 2012]{Pavlis2012-fw}
Pavlis, N.~K., Holmes, S.~A., Kenyon, S.~C., and {others} (2012).
\newblock The development and evaluation of the earth gravitational model 2008
  ({EGM2008}).
\newblock {\em research: solid earth}, 117(B4):406.

\bibitem[Penna et~al., 2008]{Penna2008-ji}
Penna, N.~T., Bos, M.~S., Baker, T.~F., and Scherneck, H.-G. (2008).
\newblock Assessing the accuracy of predicted ocean tide loading displacement
  values.
\newblock {\em J. Geodesy}, 82(12):893--907.

\bibitem[Petit and Luzum, 2010]{Petit2010-hu}
Petit, G. and Luzum, B. (2010).
\newblock {IERS} conventions (2010).

\bibitem[Poutanen et~al., 1996]{Poutanen1996-nx}
Poutanen, M., Vermeer, M., and M{\"a}kinen, J. (1996).
\newblock The permanent tide in {GPS} positioning.
\newblock {\em J. Geodesy}, 70(8):499--504.

\bibitem[Puetzfeld and L{\"a}mmerzahl, 2019]{Puetzfeld2019-sl}
Puetzfeld, D. and L{\"a}mmerzahl, C., editors (2019).
\newblock {\em Relativistic Geodesy: Foundations and Applications}.
\newblock Springer, Cham.

\bibitem[Rangelova et~al., 2016]{Rangelova2016-gj}
Rangelova, E., Sideris, M.~G., Amjadiparvar, B., and Hayden, T. (2016).
\newblock Height datum unification by means of the {GBVP} approach using tide
  gauges.
\newblock In {\em {VIII} {Hotine-Marussi} Symposium on Mathematical Geodesy},
  pages 121--129. Springer International Publishing.

\bibitem[Rawer et~al., 1978]{Rawer1978-fl}
Rawer, K., Bilitza, D., and Ramakrishnan, S. (1978).
\newblock Goals and status of the international reference ionosphere.
\newblock {\em Rev. Geophys.}, 16(2):177.

\bibitem[Riehle, 2017]{Riehle2017-yd}
Riehle, F. (2017).
\newblock Optical clock networks.
\newblock {\em Nat. Photonics}, 11:25.

\bibitem[Rummel and Teunissen, 1988]{Rummel1988-qo}
Rummel, R. and Teunissen, P. (1988).
\newblock Height datum definition, height datum connection and the role of the
  geodetic boundary value problem.
\newblock {\em Bull. Am. Assoc. Hist. Nurs.}, 62(4):477--498.

\bibitem[Sanchez, 2007]{Sanchez2007-ac}
Sanchez, L. (2007).
\newblock Definition and realisation of the {SIRGAS} vertical reference system
  within a globally unified height system.
\newblock In Tregoning, P. and Rizos, C., editors, {\em Dynamic Planet:
  Monitoring and Understanding a Dynamic Planet with Geodetic and Oceanographic
  Tools {IAG} Symposium Cairns, Australia 22--26 August, 2005}, pages 638--645.
  Springer Berlin Heidelberg, Berlin, Heidelberg.

\bibitem[S{\'a}nchez et~al., 2016]{Sanchez2016-nd}
S{\'a}nchez, L., {\v C}underl{\'\i}k, R., Dayoub, N., Mikula, K.,
  Minarechov{\'a}, Z., {\v S}{\'\i}ma, Z., Vatrt, V., and Vojt{\'\i}{\v
  s}kov{\'a}, M. (2016).
\newblock A conventional value for the geoid reference potential {0W0}.
\newblock {\em J. Geodesy}, 90(9):815--835.

\bibitem[S{\'a}nchez and Sideris, 2017]{Sanchez2017-zl}
S{\'a}nchez, L. and Sideris, M.~G. (2017).
\newblock Vertical datum unification for the international height reference
  system ({IHRS}).
\newblock {\em Geophys. J. Int.}, 209(2):570--586.

\bibitem[Sharifi et~al., 2013]{Sharifi2013-bm}
Sharifi, M.~A., Seif, M.~R., and Hadi, M.~A. (2013).
\newblock A comparison between numerical differentiation and kalman filtering
  for a leo satellite velocity determination.
\newblock {\em Artificial Satellites}, 48(3):103--110.

\bibitem[Shen et~al., 1993]{Shen1993-rb}
Shen, W., Chao, D., and Jin, B. (1993).
\newblock On relativistic geoid.
\newblock {\em Bollettino di geodesia e scienze affini}, 52(3):207--216.

\bibitem[Shen et~al., 2019]{Shen2019-od}
Shen, Z., Shen, W.-B., Peng, Z., Liu, T., Zhang, S., and Chao, D. (2019).
\newblock Formulation of determining the gravity potential difference using
  {Ultra-High} precise clocks via optical fiber frequency transfer technique.
\newblock {\em J. Earth Sci.}, 30(2):422--428.

\bibitem[Shen et~al., 2016]{Shen2016-lc}
Shen, Z., Shen, W.-B., and Zhang, S. (2016).
\newblock Formulation of geopotential difference determination using
  optical-atomic clocks onboard satellites and on ground based on doppler
  cancellation system.
\newblock {\em Geophys. J. Int.}, 206(2):1162--1168.

\bibitem[Shen et~al., 2017]{Shen2017-kg}
Shen, Z., Shen, W.-B., and Zhang, S. (2017).
\newblock Determination of gravitational potential at ground using
  {Optical-Atomic} clocks on board satellites and on ground stations and
  relevant simulation experiments.
\newblock {\em Surv. Geophys.}, 38(4):757--780.

\bibitem[Sideris, 2015]{Sideris2015-fn}
Sideris, M. (2015).
\newblock Geodetic world height system unification.
\newblock In Freeden, W., Zuhair~Nashed, M., and Sonar, T., editors, {\em
  Handbook of Geomathematics}, pages 3067--3085. Springer Berlin Heidelberg.

\bibitem[St{\"o}cker-Meier, 1990]{Stocker-Meier1990-kj}
St{\"o}cker-Meier, E. (1990).
\newblock Theory of oceanic levelling for improving the geoid from satellite
  altimetry.
\newblock {\em Bull. Am. Assoc. Hist. Nurs.}, 64(3):247--258.

\bibitem[Takano et~al., 2016]{Takano2016-dr}
Takano, T., Takamoto, M., Ushijima, I., Ohmae, N., Akatsuka, T., Yamaguchi, A.,
  Kuroishi, Y., Munekane, H., Miyahara, B., and Katori, H. (2016).
\newblock Geopotential measurements with synchronously linked optical lattice
  clocks.
\newblock {\em Nat. Photonics}, 10(10):662.

\bibitem[Torge and M{\"u}ller, 2012]{Torge2012-xq}
Torge, W. and M{\"u}ller, J. (2012).
\newblock {\em Geodesy}.
\newblock Walter de Gruyter.

\bibitem[Van~Camp and Vauterin, 2005]{Van_Camp2005-qt}
Van~Camp, M. and Vauterin, P. (2005).
\newblock Tsoft: graphical and interactive software for the analysis of time
  series and earth tides.
\newblock {\em Comput. Geosci.}, 31(5):631--640.

\bibitem[Vermeer, 1983]{Vermeer1983-lh}
Vermeer, M. (1983).
\newblock {\em Chronometric Levelling}.
\newblock Geodeettinen Laitos, Geodetiska Institutet.

\bibitem[Vessot and Levine, 1979]{Vessot1979-ot}
Vessot, R. F.~C. and Levine, M.~W. (1979).
\newblock A test of the equivalence principle using a space-borne clock.
\newblock {\em Gen. Relat. Grav.}, 10(3):181--204.

\bibitem[Vessot et~al., 1980]{Vessot1980-ax}
Vessot, R. F.~C., Levine, M.~W., Mattison, E.~M., Blomberg, E.~L., Hoffman,
  T.~E., Nystrom, G.~U., Farrel, B.~F., Decher, R., Eby, P.~B., Baugher, C.~R.,
  Watts, J.~W., Teuber, D.~L., and Wills, F.~D. (1980).
\newblock Test of relativistic gravitation with a {Space-Borne} hydrogen maser.
\newblock {\em Phys. Rev. Lett.}, 45(26):2081--2084.

\bibitem[Voigt et~al., 2017]{Voigt2017-qc}
Voigt, C., F{\"o}rste, C., Wziontek, H., Crossley, D., Meurers, B.,
  P{\'a}link{\'a}{\v s}, V., Hinderer, J., Boy, J.-P., Barriot, J.-P., and Sun,
  H. (2017).
\newblock The data base of the international geodynamics and earth tide service
  ({IGETS}).
\newblock In {\em {EGU} General Assembly Conference Abstracts}, volume~19, page
  4947.

\bibitem[Weinberg, 1972]{Weinberg1972-le}
Weinberg, S. (1972).
\newblock {\em Gravitation and Cosmology: Principles and Applications of the
  General Theory of Relativity}.
\newblock Wiley, New York.

\bibitem[Wenzel, 1996]{Wenzel1996-tk}
Wenzel, H.-G. (1996).
\newblock The nanogal software: Earth tide data processing package {ETERNA}
  3.30.
\newblock {\em Bull. Inf. Mar{\'e}es Terrestres}, 124:9425--9439.

\bibitem[Woodworth et~al., 2012]{Woodworth2012-th}
Woodworth, P.~L., Hughes, C.~W., Bingham, R.~J., and Gruber, T. (2012).
\newblock Towards worldwide height system unification using ocean information.
\newblock {\em Journal of Geodetic Science}, 2(4):302--318.

\bibitem[Wu et~al., 2019]{Wu2019-wq}
Wu, H., M{\"u}ller, J., and L{\"a}mmerzahl, C. (2019).
\newblock Clock networks for height system unification: a simulation study.
\newblock {\em Geophys. J. Int.}, 216(3):1594--1607.

\end{thebibliography}

\end{document}